\documentclass[aps,prl,reprint,twocolumn,amsmath,amssymb,floatfix,nofootinbib,superscriptaddress]{revtex4-2}
\usepackage{amsmath,amssymb,amsthm,booktabs,graphicx,hyperref}
\usepackage{tikz}
\usetikzlibrary{arrows.meta,calc,positioning,decorations.pathreplacing}
\DeclareMathVersion{figsans}
\SetSymbolFont{operators}{figsans}{OT1}{phv}{m}{n}
\SetSymbolFont{letters}{figsans}{OML}{ntxsfmi}{m}{it}
\SetSymbolFont{symbols}{figsans}{OMS}{cmsy}{m}{n}
\SetMathAlphabet{\mathrm}{figsans}{OT1}{phv}{m}{n}
\SetMathAlphabet{\mathbf}{figsans}{OT1}{phv}{b}{n}
\hypersetup{hidelinks}
\newtheorem{theorem}{Theorem}
\newtheorem{lemma}[theorem]{Lemma}
\newtheorem{proposition}[theorem]{Proposition}

\newcommand{\F}{\mathbb F_2}
\newcommand{\CCZ}{\mathrm{CCZ}}
\newcommand{\CCCZ}{\mathrm{CCCZ}}
\newcommand{\CS}{\mathrm{CS}}
\newcommand{\tr}{\operatorname{Tr}}
\newcommand{\RR}{\raggedright}
\newcommand{\TNmax}{8}
\newcommand{\TLB}{9}
\newcommand{\QuarticUB}{48}
\newcommand{\QuarticCoef}{864}
\begin{document}
\title{The limits of magic-state distillation}
\author{Lloren\c{c} Balada Gaggioli}
\affiliation{Czech Technical University in Prague, Prague, Czech Republic}
\affiliation{LAAS-CNRS, Universit\'e de Toulouse, France}
\author{Jakub Mare\v{c}ek}
\affiliation{Czech Technical University in Prague, Prague, Czech Republic}
\date{\today}
\begin{abstract}
Magic-state factories are a major cost of fault-tolerant quantum computing, yet how few noisy inputs they need is known only for restricted protocols. We show that, for tolerating input faults, any stabilizer protocol, however adaptive, postselected or multistage, is no better than a projection onto one stabilizer code, and that for rational targets a fault pattern is harmful exactly when its complement is. Searches over all codes show that quadratic distillation of one T state needs exactly nine inputs and that the eight-input CCZ factory is optimal. At order $d$, one T state needs $n\geq2d$.
\end{abstract}
\maketitle

\section{Introduction}
Fault-tolerant quantum computers implement Clifford operations cheaply, but universal computation also needs non-Clifford resources, usually supplied as magic states prepared by distillation~\cite{bravyikitaev,bravyihaah} or cultivation~\cite{cultivation,googlecultivation}. Magic-state factories are a large part of the overhead of fault-tolerant algorithms, and much work has gone into making them cheaper~\cite{litinski,gidneyfowler,campbellhoward,eastin,jones,haah,borrowed,jacinto,wjs,gong}. How far can they improve? We ask how many noisy T states a protocol must consume to output a target state with error $O(p^d)$, where $p$ is the input error rate and $d$ the \emph{suppression order}. Small $d$ matters most in the near term, because cultivated inputs already have low error.

\begin{figure*}[t]
\centering
\definecolor{tolblue}{cmyk}{1,0.36,0,0.30}      
\definecolor{harmver}{cmyk}{0,0.56,1,0.165}     
\definecolor{skyblue}{cmyk}{0.63,0.23,0,0.09}   
\definecolor{rejgray}{cmyk}{0,0,0,0.62}         
\definecolor{inkgray}{cmyk}{0,0,0,0.75}         
\newcommand{\figPi}{\text{\fontencoding{OT1}\fontfamily{cmss}\selectfont\char5}}
\begin{tikzpicture}[
  font=\fontsize{8}{9.6}\fontfamily{phv}\selectfont\mathversion{figsans},
  line width=0.5pt,
  >={Stealth[length=1.9mm,width=1.4mm]},
  letter/.style={font=\fontsize{8}{9.6}\fontfamily{phv}\fontseries{b}\selectfont, anchor=north west, inner sep=0pt},
  gate/.style={draw, rounded corners=1pt, fill=black!8},
  code/.style={draw=skyblue, rounded corners=1pt, fill=skyblue!18},
  cl/.style={double=white, double distance=0.9pt, line width=0.4pt, rejgray},
  brace/.style={decorate, decoration={brace, amplitude=2.2pt}},
  meter/.pic={
    \draw[fill=white] (-0.15,-0.12) rectangle (0.15,0.12);
    \draw (-0.095,-0.055) arc (160:20:0.1);
    \draw[->, >={Stealth[length=0.9mm,width=0.8mm]}] (0,-0.07) -- (0.08,0.08);
  },
  vec/.style={->, line width=0.8pt},
  drop/.style={dotted, line width=0.6pt},
  proj/.style={->, line width=1pt},
]
\begin{scope}[shift={(0,0)}]
  \node[letter] at (0,4.5) {(a)};
  \draw[brace] (0.92,2.83) -- (0.92,4.05);
  \node[anchor=east] at (0.86,3.44) {$\rho_T^{\otimes n}$};
  \draw (1.0,3.95) -- (4.05,3.95);
  \draw (1.0,3.61) -- (4.05,3.61);
  \draw (1.0,3.27) -- (2.95,3.27);
  \draw (1.0,2.93) -- (1.95,2.93);
  \draw[gate] (1.22,2.80) rectangle (1.55,4.08);
  \draw[cl] (1.95,3.05) -- (1.95,3.44) -- (2.30,3.44);
  \pic at (1.95,2.93) {meter};
  \draw[gate] (2.30,3.14) rectangle (2.60,4.08);
  \draw[cl] (2.95,3.39) -- (2.95,3.78) -- (3.30,3.78);
  \pic at (2.95,3.27) {meter};
  \draw[gate] (3.30,3.48) rectangle (3.60,4.08);
  \draw[brace] (4.10,4.03) -- (4.10,3.53);
  \node[anchor=west] at (4.17,3.78) {$\tau$};
  \draw[->, line width=0.8pt, inkgray] (2.30,2.62) -- (2.30,2.12);
  \node[anchor=west, align=left, inkgray] at (2.42,2.37) {one accepted\\branch};
  \draw[brace] (0.92,0.63) -- (0.92,1.85);
  \node[anchor=east] at (0.86,1.24) {$\rho_T^{\otimes n}$};
  \foreach \y in {1.75,1.41,1.07,0.73} \draw (1.0,\y) -- (2.30,\y);
  \draw (2.80,1.75) -- (4.05,1.75);
  \draw (2.80,1.41) -- (4.05,1.41);
  \draw[code] (1.22,0.60) rectangle (1.88,1.88);
  \node at (1.55,1.24) {$\figPi_C$};
  \draw[gate] (2.30,0.60) rectangle (2.80,1.88);
  \node at (2.55,1.24) {$V$};
  \draw[brace] (4.10,1.83) -- (4.10,1.33);
  \node[anchor=west] at (4.17,1.58) {$\tau$};
\end{scope}
\begin{scope}[shift={(5.15,0)}]
  \node[letter] at (0,4.5) {(b)};
  \coordinate (O) at (2.55,0.58);
  \coordinate (t) at (1,0.17);                       
  \fill[skyblue!15] (0.15,0.15) -- (4.75,0.15) -- (6.25,1.20) -- (1.65,1.20) -- cycle;
  \draw[skyblue] (0.15,0.15) -- (4.75,0.15) -- (6.25,1.20) -- (1.65,1.20) -- cycle;
  \node[skyblue!70!black] at (4.45,0.32) {$C$};
  \draw[tolblue!40, line width=1.8pt, line cap=round] ($(O)-2.0*(t)$) -- ($(O)+3.3*(t)$);
  \node[anchor=south west, tolblue, inner sep=1pt] at ($(O)+3.3*(t)+(0.02,0.04)$) {$\mathbb C\,U|\tau\rangle$};
  \coordinate (f0) at ($(O)+2.85*(t)$);  \coordinate (p0) at ($(f0)+(0,2.30)$);
  \coordinate (f1) at ($(O)-1.45*(t)$);  \coordinate (p1) at ($(f1)+(0,2.55)$);
  \coordinate (fh) at ($(O)+(0.80,-0.30)$); \coordinate (ph) at ($(fh)+(0,3.00)$);
  \coordinate (pr) at ($(O)+(0,2.97)$);
  \draw[tolblue, line width=0.4pt] ($(f0)-0.13*(t)$) -- ++(0,0.13) -- ($(f0)+(0,0.13)$);
  \draw[tolblue, line width=0.4pt] ($(f1)+0.13*(t)$) -- ++(0,0.13) -- ($(f1)+(0,0.13)$);
  \draw[harmver, line width=0.4pt] ($(fh)+(-0.122,0.046)$) -- ++(0,0.13) -- ($(fh)+(0,0.13)$);
  \draw[rejgray, line width=0.4pt] ($(O)+(0.106,0.074)$) -- ++(0,0.13) -- ($(O)+(0,0.13)$); 
  \draw[drop, tolblue] (p0) -- (f0);
  \draw[drop, tolblue] (p1) -- (f1);
  \draw[drop, harmver] (ph) -- (fh);
  \draw[proj, tolblue] (O) -- (f0);
  \draw[proj, tolblue] (O) -- (f1);
  \draw[proj, harmver] (O) -- (fh);
  \draw[vec, rejgray] (O) -- (pr);
  \draw[vec, harmver] (O) -- (ph);
  \draw[vec, tolblue] (O) -- (p0);
  \draw[vec, tolblue] (O) -- (p1);
  \fill (O) circle (0.9pt);
  \node[anchor=south, rejgray, inner sep=1.5pt, align=center] at (pr) {rejected\\[-1pt]$\figPi_C\phi_e=0$};
  \node[anchor=west, harmver, inner sep=1.5pt] at (ph) {harmful};
  \node[anchor=west, tolblue, inner sep=1.5pt] at (p0) {$\phi_{\mathbf 0}$};
  \node[anchor=east, tolblue, inner sep=1.5pt] at (p1) {$\phi_{\mathbf 1}$};
\end{scope}
\begin{scope}[shift={(12.25,0)}]
  \node[letter] at (0,4.5) {(c)};
  \fill[tolblue!14, rounded corners=2pt] (0.08,1.27) rectangle (1.07,1.63);
  \fill[harmver!14, rounded corners=2pt] (1.18,1.27) rectangle (3.82,1.63);
  \fill[tolblue!14, rounded corners=2pt] (3.93,1.27) rectangle (4.92,1.63);
  \draw[inkgray, dashed, line width=0.5pt] (2.50,1.20) -- (2.50,3.92);
  \node[inkgray] at (2.50,4.18) {$\sigma\colon\ \sqrt2\mapsto-\sqrt2,\quad p\mapsto1-p$};
  \foreach \w/\h in {0/2.55,1/1.95,2/1.35,3/0.75} {
    \pgfmathsetmacro{\xa}{0.30+0.55*\w}
    \pgfmathsetmacro{\xb}{0.30+0.55*(8-\w)}
    \draw[<->, inkgray, >={Stealth[length=1.5mm,width=1.1mm]}]
      (\xa,1.62) .. controls (\xa,1.62+\h) and (\xb,1.62+\h) .. (\xb,1.62);
  }
  \foreach \w in {0,...,8} {
    \pgfmathsetmacro{\x}{0.30+0.55*\w}
    \fill (\x,1.45) circle (1.2pt);
    \node at (\x,1.06) {\w};
  }
  \node[tolblue] at (0.575,0.68) {harmless};
  \node[harmver] at (2.50,0.68) {may damage};
  \node[tolblue] at (4.425,0.68) {harmless};
  \node at (2.50,0.27) {number of faults $|e|$};
\end{scope}
\end{tikzpicture}
\caption{Distillation protocols as stabilizer codes. A protocol takes $n$ noisy T states $\rho_T(p)$, each faulty ($Z|T\rangle$ in place of $|T\rangle$) with probability $p$, and outputs a target state $\tau$. A fault pattern $e\in\F^n$ marks the faulty inputs, and the corresponding pure inputs $\phi_e=Z^e|T\rangle^{\otimes n}$ are orthonormal. A pattern is harmless if, on that input, the protocol outputs exactly $\tau$ whenever it accepts, and harmful otherwise. The protocol has suppression order $d$, meaning an output error $O(p^d)$, when it accepts the fault-free input and all patterns with fewer than $d$ faults are harmless. (a)~Any protocol, whatever its Clifford gates, Pauli measurements, feedforward and postselection (top), can be replaced by one of its accepted branches without loss of suppression order (Theorem~\ref{thm:code}). That branch measures the stabilizers of a single code $C$, keeps one outcome, which applies the projector $\Pi_C$, and decodes with a Clifford $V$ (bottom). (b)~Under the assumptions of Theorem~\ref{thm:code}, pattern $e$ is harmless exactly when $\Pi_C$ (dotted lines) maps $\phi_e$ onto the line $\mathbb C\,U|\tau\rangle$ spanned by the encoded target, as for $\phi_{\mathbf0}$ and $\phi_{\mathbf1}$, or to zero, in which case $\phi_e$ is orthogonal to $C$ and the run is always rejected. A harmful pattern lands elsewhere in $C$. The vectors are drawn schematically. (c)~If all Pauli expectations of $\tau$ are rational, as for CCZ, the automorphism $\sigma\colon\sqrt2\mapsto-\sqrt2$ commutes with stabilizer operations and maps $\rho_T(p)$ to $\rho_T(1-p)$. A pattern is then harmless exactly when its complement is (Theorem~\ref{thm:pairing}), so harmlessness is symmetric under reflection of the number of faults $|e|$ about $n/2$. For $n=8$ and $d=2$, the harmless weights 0 and 1 are mirrored at 8 and 7, and only patterns with two to six faults can be harmful. Section~\ref{app:example} works through explicit examples.}
\label{fig:reductions}
\end{figure*}
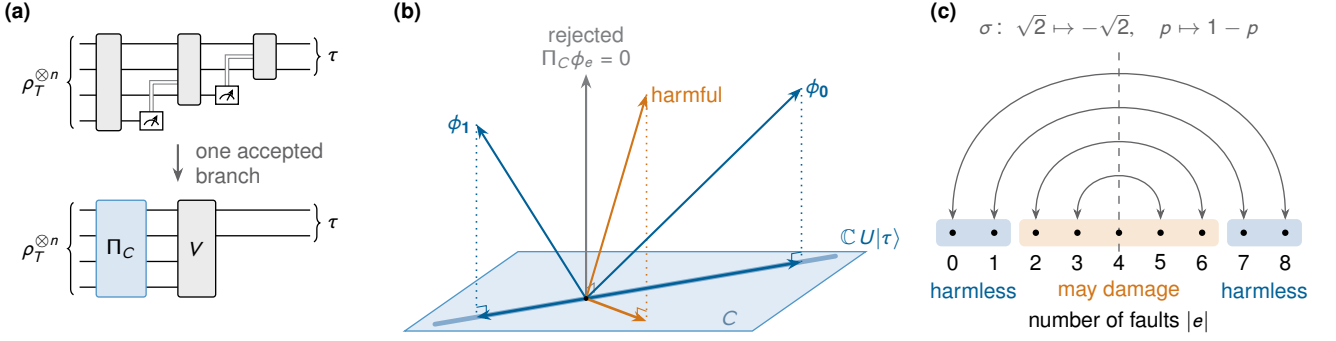

Earlier work answers parts of this question. Resource theories bound all protocols at once, but the resource measures that remain valid under postselection grow only like $1/p$ for a noisy T state and give little more than $n\gtrsim d$~\cite{fangliu,fangliu2,regulaprl}. Stronger bounds assume that the protocol is deterministic or succeeds with high probability~\cite{regulatakagi,seddon}, whereas ours hold however rarely it accepts. Exact conversions without noise are well understood, from the canonical form of postselected stabilizer circuits~\cite{beverland} to the recent result that a universal protocol making one CCZ state from copies of an unknown pure magic qubit state needs exactly six copies~\cite{rizzoleone}, but they say nothing about suppression order. Exhaustive searches and classifications resolve $d$ precisely, yet only within particular code families or circuit layouts~\cite{nezamihaah,wjs,jacinto,borrowed,kbk}. Even for the CCZ state, whose eight-input quadratic factory~\cite{eastin,jones,haah} underlies many resource estimates~\cite{litinski,gidneyfowler}, it was not known whether an adaptive or multistage protocol could use fewer inputs.

For small factories, we turn the question into a combinatorial one with two reductions. First, adaptivity, randomness, postselection and multiple stages do not help a protocol tolerate input faults; this extends the normal forms of Campbell and Browne~\cite{campbellbrowne} and of Beverland \emph{et al.}~\cite{beverland}. A set of fault patterns can be tolerated if and only if some stabilizer code on the inputs projects each of them onto one common vector, which encodes the target, or to zero. Second, flipping the sign of $\sqrt2$ commutes with stabilizer operations and fixes every target with rational Pauli expectations, CCZ among them, but turns a T state with error $p$ into one with error $1-p$, so a fault pattern damages the output exactly when its complement does. The first reduction replaces an unbounded space of protocols by the stabilizer codes on the $n$ inputs, and the second enlarges the set of patterns that must be tolerated, which eliminates many candidate codes. With symmetry reduction and exact arithmetic, we can then search all codes, up to $9\times10^{11}$ signed codes in the largest run. Analytic arguments extend the lower bounds to every order. Integer and satisfiability programs serve only to find explicit factories, which we certify in exact arithmetic.

Together, these methods give the first exact minimum input counts for error suppression by arbitrary stabilizer protocols (Table~\ref{tab:final}). At $d=2$, one T state needs nine noisy inputs: no protocol works with eight, and the search found a nine-input factory with ideal-input acceptance probability $1/128$, using one fewer input than the previous best. The controlled-S (CS) state needs eight, and no adaptive, postselected or multistage protocol beats the eight-input CCZ factory. In all, lower and upper bounds meet in seven entries of Table~\ref{tab:final}. At every order $d$, rational targets other than stabilizer states need $n\geq2d+1$ inputs and one T state needs $n\geq2d$. To our knowledge, these are the first bounds that survive arbitrary postselection and exceed $\max(d,\nu)$, where $\nu$ is the stabilizer nullity of the target. On the constructive side, a two-level design gives a quartic CCZ factory with \QuarticUB\ inputs and less than a third of the previous leading error.

\section{Factories as codes}\label{sec:framework}
\paragraph*{Setting.}
Let $|T\rangle=(|0\rangle+\omega|1\rangle)/\sqrt2$ with $\omega=e^{i\pi/4}$, Clifford-equivalent to the H-type state of~\cite{bravyikitaev} (not to their T-type state, which the five-qubit code distills). After the standard twirl~\cite{bravyikitaev}, each input is
\begin{equation}
 \rho_T(p)=(1-p)|T\rangle\langle T|+pZ|T\rangle\langle T|Z=\tfrac12\bigl[I+\tfrac{1-2p}{\sqrt2}(X+Y)\bigr].
 \label{eq:input}
\end{equation}
Clifford operations are assumed ideal, and we count the noisy inputs consumed per run. A fault pattern $e\in\F^n$ marks the faulty inputs, the states $|\phi_e\rangle=Z^e|T\rangle^{\otimes n}$ are orthonormal, and $\rho_T(p)^{\otimes n}=\sum_ew_e\rho_e$ with $\rho_e=|\phi_e\rangle\langle\phi_e|$ and $w_e=p^{|e|}(1-p)^{n-|e|}$. A \emph{protocol} applies stabilizer operations to the $n$ inputs: Clifford gates, stabilizer ancillas, Pauli measurements, classical feedforward and randomness, postselection and discarding, in any number of stages. Its accepted output is a trace-nonincreasing map $\Lambda$, which does not depend on $p$. For a pure target $\tau$ of $r$ qubits with projector $\Pi$, pattern $e$ has \emph{damage} $h_e=\tr[(I-\Pi)\Lambda(\rho_e)]\geq0$ and is \emph{harmless} if $h_e=0$, that is, if $\Lambda(\rho_e)\propto\Pi$. The output error is $\epsilon=\sum_ew_eh_e/\sum_ew_e\tr\Lambda(\rho_e)$, so a protocol that accepts the fault-free pattern has suppression order $d$ ($\epsilon=O(p^d)$) exactly when every pattern with $|e|<d$ is harmless, and every order-$d$ protocol can be brought into this form (Lemma~\ref{lem:clean}). Theorems~\ref{thm:code} and~\ref{thm:pairing} concern individual patterns and hold for any fault distribution, whereas suppression orders and error coefficients refer to independent faults. We write $n_{\min}(\tau,d)$ for the least possible $n$, which is nondecreasing in $d$. The \emph{stabilizer nullity} $\nu(\tau)$ is $r$ minus the number of independent Pauli operators that fix $\tau$~\cite{beverland}, for example $\nu=3$ for CCZ. A target is \emph{rational} (\emph{dyadic}) if all its Pauli expectations are rational (of the form $k/2^j$). Phase states $2^{-r/2}\sum_xi^{f(x)}|x\rangle$ with integer $f$, such as CCZ, CS and CCCZ, are dyadic, whereas $|T\rangle$ is not rational.

The first reduction replaces every protocol by a single code projection.
\begin{theorem}\label{thm:code}
Let $\tau$ be fixed by no Pauli operator other than the identity, that is, $\nu(\tau)=r$, and let $H$ be a set of fault patterns containing $\mathbf 0$. Some protocol accepts the fault-free input and makes every pattern in $H$ harmless if and only if there is a stabilizer code $C$ on the $n$ inputs, with a Clifford encoding $U$ of $r$ qubits, such that $\Pi_C|\phi_{\mathbf0}\rangle\neq0$ and
\begin{equation}
 \Pi_C|\phi_e\rangle\in\mathbb C\,U|\tau\rangle\qquad\text{for every }e\in H.
 \label{eq:collapse}
\end{equation}
\end{theorem}
Figure~\ref{fig:reductions} illustrates this in panels (a) and (b). Campbell and Browne showed that the fidelity of any single-output protocol is bounded by that of a code projection~\cite{campbellbrowne}; our proof (Sec.~\ref{app:proofs}) refines the protocol in the same way, for all fault patterns at once. Since damages are nonnegative, every branch is harmless on $H$, and each Kraus operator is a code projection followed by a Clifford isometry~\cite{fattal}. We call Eq.~\eqref{eq:collapse} \emph{collapse}: every tolerated input is projected onto one code vector or is rejected ($\Pi_C|\phi_e\rangle=0$). With the Gram matrix $G_{ef}=\langle\phi_e|\Pi_C|\phi_f\rangle$, and given that $\Pi_C|\phi_{\mathbf0}\rangle$ encodes $\tau$, collapse is the set of polynomial conditions $G_{\mathbf0\mathbf0}>0$ and $G_{\mathbf0\mathbf0}G_{ee}=|G_{\mathbf0e}|^2$ for $e\in H$; rejected patterns have $G_{ee}=G_{\mathbf0e}=0$.

\begin{table*}[t]
\caption{Fewest noisy T inputs $n_{\min}(\tau,d)$ for distilling target $\tau$ with suppression order $d$, over all stabilizer protocols, including adaptive, postselected and multistage ones. The column $n_{\min}$ gives the exact value where the lower and upper bounds meet. ``Prior LB'' is $\max(d,\nu)$~\cite{fangliu,fangliu2,regulaprl,beverland}, the best general bound we found stated under arbitrary postselection (deterministic protocols obey stronger bounds~\cite{regulatakagi}; Sec.~\ref{app:prior}). Bold entries are new. The last column lists upper bounds explicitly reported before. Entries marked $\dagger$ follow immediately from earlier results but are not stated there: the $d=1$ lower bounds from the canonical form of~\cite{beverland} (Theorem~\ref{thm:analytic}), the CS factory from~\cite[Proposition~A.4]{beverland} and the eight-input CCZ factory, and the 36-input CCZ factory from~\cite{wjs,howardcampbell}. Lower bounds extend to higher $d$ by Lemma~\ref{lem:trade}. Sources: $a$, analytic (Theorem~\ref{thm:analytic}); $\ell$, Lemma~\ref{lem:trade} applied to a search result; $e$, exhaustive search (Propositions~\ref{thm:tcs} and~\ref{thm:ccz}); $c$, Clifford orbit (Proposition~\ref{prop:orbit}); $m$, integer program, certified; $s$, CCZ factory followed by a Y measurement (Proposition~\ref{prop:cs}); $p$, explicit code projection (Proposition~\ref{prop:cccz}); $t$, two-level design, certified; $x$, found by the exhaustive search and certified (Proposition~\ref{prop:t9}). Leading error coefficients are in parentheses.}
\label{tab:final}
\begin{ruledtabular}
\begin{tabular}{ccccccc}
$\tau$ & $d$ & $n_{\min}$ & prior LB & LB & UB & previously reported UB \\
\hline
T & 2 & $\mathbf{9}$ & 2 & $\mathbf{\TLB}^e$ & $\mathbf{9}$ $(\tfrac{27}{2})^x$ & 10~\cite{mek} \\
T & 3 & & 3 & $\mathbf{10}^\ell$ & 15 & 15 (35)~\cite{bravyikitaev} \\
T & 4 & & 4 & $\mathbf{11}^\ell$ & 48 & 48~\cite{wjs} \\
T & 5 & & 5 & $\mathbf{12}^\ell$ & 49 (1411) & 49 (1411)~\cite{bravyihaah} \\
T & $d\geq6$ & & $d$ & $\mathbf{\max(2d,d{+}7)}^{a,\ell}$ & -- & -- \\
CS & 1 & 3 & 2 & 3$^{a\dagger}$ & 3 & 3 \\
CS & 2 & $\mathbf{8}$ & 2 & $\mathbf{8}^e$ & $8^{s\dagger}$ & 12~\cite{campbellhoward,gong} \\
CS & 3 & & 3 & $\mathbf{9}^{e,\ell}$ & 35 & 35~\cite{nezamihaah,wjs} \\
CCZ & 1 & 4 & 3 & 4$^{a\dagger}$ & 4 & 4~\cite{jones} \\
CCZ & 2 & $\mathbf{8}$ & 3 & $\mathbf{8}^e$ & 8 (28) & 8 (28)~\cite{eastin,jones,haah} \\
CCZ & 3 & & 3 & $\mathbf{10}^e$ & 36 (120)$^{c,m\dagger}$ & 47 (236)~\cite{jacinto,wjs} \\
CCZ & 4 & & 4 & $\mathbf{11}^\ell$ & \QuarticUB\ ($\mathbf{\QuarticCoef}$)$^t$ & 48 (2816)~\cite{jacinto} \\
CCZ & $d\geq5$ & & $d$ & $\mathbf{\max(2d{+}1,d{+}7)}^{a,\ell}$ & $O(d)$~\cite{golowich,nguyen} & -- \\
CCZ$^{\otimes2}$ & 1 & $\mathbf{8}$ & 6 & $\mathbf{8}^e$ & 8 & 8~\cite{jones} \\
CCZ$^{\otimes2}$ & 2 & & 6 & $\mathbf{10}^e$ & 14 & 14~\cite{campbellhoward} \\
CCCZ & 1 & $\mathbf{6}$ & 4 & $\mathbf{6}^e$ & 6$^p$ & 6~\cite{beverland} \\
\end{tabular}
\end{ruledtabular}
\end{table*}

The second reduction concerns rational targets.
\begin{theorem}\label{thm:pairing}
If $\tau$ is rational, then in every protocol a pattern $e$ is harmless if and only if its complement $e+\mathbf1$ is. With suppression order $d$, every pattern with $|e|<d$ or $|e|>n-d$ is harmless.
\end{theorem}
Every branch of a protocol maps Pauli coefficients by rational matrices and therefore commutes with the automorphism $\sigma$ of $\mathbb Q(\sqrt2)$ that flips the sign of $\sqrt2$. By Eq.~\eqref{eq:input}, $\sigma$ maps $\rho_T(p)$ to $\rho_T(1-p)$, hence $\rho_e$ to $\rho_{e+\mathbf1}$, and it fixes a rational target. Branch by branch, the damages of $e$ and of its complement are therefore conjugate in $\mathbb Q(\sqrt2)$, and one vanishes exactly when the other does (Sec.~\ref{app:proofs}). This map is the $\sqrt2$-conjugation of exact synthesis~\cite{rossselinger}, and Beverland \emph{et al.} used it to bound postselection probabilities~\cite[Proposition~5.8]{beverland}; applied to noisy inputs, it pairs each fault pattern with its complement. In the code picture, extend $\sigma$ to amplitudes in $\mathbb Q(\omega)=\mathbb Q(\sqrt2,i)$ by fixing $i$. Then $\sigma(|T\rangle)=-Z|T\rangle$ and, since $\Pi_C$ has entries in $\mathbb Q(i)$, $\sigma(\Pi_C|\phi_{\mathbf0}\rangle)=(-1)^n\Pi_C|\phi_{\mathbf1}\rangle\neq0$. The fault-free and all-faulty inputs must therefore collapse together, and the projected ideal input is proportional to its own image under $\sigma$. We call this the \emph{cross condition} [Fig.~\ref{fig:reductions}(c)].

\paragraph*{A worked example.}
Three T states make $\CS|{+}{+}\rangle$: measure $Z_1Z_2Z_3$, keep the outcome $+1$ and decode, which succeeds with probability $1/2$. The all-faulty input collapses onto the same vector as the fault-free one, because $Z^{\otimes3}$ is the stabilizer, as Theorem~\ref{thm:pairing} requires, whereas a single fault acts as a logical Pauli operator and is harmful. The protocol thus has suppression order one, with output error $3p(1-p)$. Section~\ref{app:example} gives every step and two further examples.

\paragraph*{Targets up to Clifford equivalence.}
Appending a stabilizer branch that maps $\tau$ to $\tau'$ with nonzero probability turns every order-$d$ factory for $\tau$ into one for $\tau'$ (Lemma~\ref{lem:post}). In particular, $n_{\min}$ depends only on the Clifford orbit of the target, which is larger than the classes up to CNOT and S gates used to organize code families~\cite{wjs}. Howard and Campbell found that the CCZ and $\CS_{12}\CS_{13}$ states are Clifford-equivalent~\cite{howardcampbell}; the following circuit makes this explicit.
\begin{proposition}\label{prop:orbit}
\begin{equation}
 \CS_{12}\CS_{13}|{+}{+}{+}\rangle=\mathrm{CX}_{23}H_2\mathrm{CX}_{12}S_2H_2|\CCZ\rangle,
 \label{eq:orbit}
\end{equation}
with the rightmost gate acting first.
\end{proposition}

\section{Lower bounds}\label{sec:lower}
\paragraph*{Analytic bounds.}
For a rational target and small $n$, Theorem~\ref{thm:pairing} forces all but a narrow range of fault patterns to be harmless. We then construct a \emph{witness}, a state whose expansion in the fault basis contains only harmless patterns, and show that the protocol would turn it into the target, which stabilizer operations cannot do. The witnesses (Sec.~\ref{app:analytic}) are $|0\rangle^{\otimes n}$, a GHZ-type stabilizer state, low-nullity states with non-dyadic Pauli expectations and, for one T state, singlets combined with $|{\pm}\rangle$ states.
\begin{theorem}\label{thm:analytic}
Let $\tau$ not be a stabilizer state. Every protocol of suppression order $d\geq1$ satisfies $n\geq d-1+L(\tau)$, where $L(\tau)$ is the least number of ideal T states from which $\tau$ can be made with nonzero probability. If $\tau$ is rational, then $L(\tau)\geq\nu(\tau)+1$, so that $n\geq d+\nu(\tau)$, and $n\geq2d+1$. If $\tau$ is dyadic and $d\geq2$, then $n\geq d+\nu(\tau)+1$. For one T state and $d\geq2$, $n\geq2d$.
\end{theorem}
The bound $L(\tau)\geq\nu(\tau)+1$ is the equality case of the canonical form of Beverland \emph{et al.} for postselected stabilizer circuits~\cite{beverland}; the bounds $2d+1$ and $d+\nu+1$ rest on Theorem~\ref{thm:pairing}. For one T state, $\sigma$ instead exchanges $|T\rangle$ and $Z|T\rangle$, so the complement of a harmless pattern yields exactly $Z|T\rangle$; together with an inequality obeyed by two-qubit stabilizer states, this gives $n\geq2d$. We found this argument by reading off the certificate of a linear-programming relaxation of the set of protocols (Sec.~\ref{app:analytic}). Linear growth in $d$ is the correct scaling: asymptotically good codes with transversal CCZ~\cite{golowich,nguyen,whyamasaki}, fed with four-T CCZ preparations, give one CCZ state with $n=O(d)$.

\paragraph*{Exhaustive search.}
Theorem~\ref{thm:code} reduces the existence of a protocol to a search over signed stabilizer codes, that is, stabilizer codes together with a choice of generator signs. We parametrize each code by the span of the X parts of its stabilizers, of dimension $a$ (the \emph{X-rank}), its Z-type stabilizers and the Z parts that complete each X part to a stabilizer, and reduce modulo input permutations and the Clifford $W=(X+Y)/\sqrt2$, which fixes $|T\rangle$. A simple necessary condition eliminates most codes: no harmless $Z^e$ may act as a nontrivial logical operator (Lemma~\ref{lem:logical}). The remaining collapse identities are tested in exact arithmetic over $\mathbb Z[\sqrt2,i]$ (Sec.~\ref{app:search}).
\begin{proposition}[computer-assisted]\label{thm:tcs}
No stabilizer protocol distills one T state with suppression order two from $n\leq\TNmax$ inputs. None distills $\CS|{+}{+}\rangle$ with order two from $n\leq7$ inputs, or with order three from $n\leq8$.
\end{proposition}
\begin{proposition}[computer-assisted]\label{thm:ccz}
Let $\tau$ be a 3-qubit state fixed by no Pauli operator other than the identity, for example CCZ. No stabilizer protocol has suppression order two with $n\leq7$ inputs. If $\tau$ is also rational, none has suppression order three with $n\leq9$. Two CCZ states need eight ideal T states and, for quadratic suppression, at least ten noisy ones. The CCCZ state needs six ideal T states.
\end{proposition}
For one T state, which is not rational, the search uses only patterns of weight below $d$. At $n=8$ it finds codes that collapse, but none of them encodes a state Clifford-equivalent to $|T\rangle$; most encode stabilizer states (Sec.~\ref{app:search}). At $n=9$ it finds a T-state factory, so the bound is tight (Proposition~\ref{prop:t9}). By Proposition~\ref{thm:ccz}, the eight-input quadratic CCZ factory is optimal, and among codes of X-rank at most four it is unique up to symmetry. At $n=7$ the weight-one collapse conditions alone exclude every code, which is why the quadratic bound needs no rationality. Feeding $d-s$ inputs with free maximally mixed qubits, $I/2=\rho_T(\tfrac12)$, turns an order-$d$ protocol into an order-$s$ one, so $n_{\min}(\tau,d)\geq n_{\min}(\tau,s)+d-s$ for every non-stabilizer target (Lemma~\ref{lem:trade}), and the searches give $n_{\min}(T,d)\geq d+7$ for $d\geq2$ and $n_{\min}(\CCZ,d)\geq d+7$ for $d\geq3$.

\section{Upper bounds and exact values}\label{sec:upper}
Explicit factories give the upper bounds, and where they meet the lower bounds above they fix $n_{\min}$ exactly: for one T state, CS and CCZ at $d=2$, and at $d=1$, where $n_{\min}(\tau,1)=L(\tau)$, for CS, CCZ, two CCZ states and CCCZ. We certify every construction below by the collapse test of Theorem~\ref{thm:code}, in exact arithmetic.

\paragraph*{A nine-input T factory from the search.}
A code that the exhaustive search finds to collapse onto an encoded target is itself a factory. For one T state at $n=9$ and $d=2$, the codes of X-rank at most six include 96 such codes, all equivalent to the following one.
\begin{proposition}\label{prop:t9}
Take nine noisy T states, measure the eight stabilizers
\begin{equation*}
\begin{array}{ll}
-\mathrm{XZIIZIIIX}, & +\mathrm{XIZIZIIXI},\\
+\mathrm{ZXZZIIXZI}, & +\mathrm{ZZXIZXIIZ},\\
-\mathrm{YXYIXIIZZ}, & -\mathrm{YYXXIIIII},\\
+\mathrm{IZIZZIZII}, & +\mathrm{IIZZZZIII},
\end{array}
\end{equation*}
and postselect on all outcomes $+1$. This succeeds with probability $1/128$ at $p=0$, and after a Clifford decoding the output is $|T\rangle$ with error $\epsilon=\tfrac{27}{2}p^2+O(p^3)$. Hence $n_{\min}(T,2)=9$.
\end{proposition}
An exact certificate (Sec.~\ref{app:search}) confirms that all nine single faults collapse and that the encoded state has logical Pauli expectations of moduli $\{0,1/\sqrt2,1/\sqrt2\}$, as $|T\rangle$ does. The best previous quadratic protocol uses ten inputs and outputs two T states~\cite{mek}. The new code is not of generalized triorthogonal form (Sec.~\ref{app:programs}).

\paragraph*{An eight-input CS factory by post-processing.}
Measuring one qubit of a CCZ state in the Y basis leaves $\CS|{+}{+}\rangle$ up to a Clifford correction~\cite[Proposition~A.4]{beverland} (Proposition~\ref{prop:cs}). The eight-input CCZ factory followed by this measurement is therefore a quadratic CS factory with eight inputs. With Proposition~\ref{thm:tcs} this gives $n_{\min}(\CS,2)=8$.

\paragraph*{A 36-input cubic CCZ factory.}
Many factories are built from a binary matrix with one column per input and a set of logical and check rows, subject to parity conditions (\emph{generalized triorthogonality})~\cite{bravyihaah,haah,wjs}. The classification of Wills, Jain and Singh~\cite{wjs} identifies outputs only up to CNOT and S gates, and its CCZ class needs 47 inputs at $d=3$. Its $\CS_{12}\CS_{13}$ class reaches distance three with 36 inputs and, by Proposition~\ref{prop:orbit}, an equivalence that~\cite{wjs} recalls from~\cite{howardcampbell}, it also yields CCZ states. Because the conditions are parities of products of binary variables, a short integer program rediscovers such a matrix (Sec.~\ref{app:programs}). The exact test certifies order three with leading error $120p^3$, as in~\cite{wjs}.

\paragraph*{Two-level factories.}
Among generalized triorthogonal codes, none uses fewer than 36 inputs for cubic or 48 for quartic CCZ~\cite{wjs}. Since any composition of stabilizer operations is again a protocol, an outer code may consume, besides raw T states, CCZ states produced by small inner factories. We use two inner factories: Jones' four-T preparation, which a single fault spoils, and the eight-input CCZ factory (the cube factory), which only two faults spoil. Each block applies a CCZ gate to three parities of the outer registers by gate teleportation~\cite{gottesmanchuang}, and a faulty block applies a Z error to a nonzero combination of them. The composite has suppression order $D$ if every combination of faulty units whose total cost (one for a T input or Jones block, two for a cube block) is below $D$ is detected or harmless. All conditions remain parities, now including the cubic terms contributed by the blocks, and a satisfiability (SAT) solver finds designs quickly (Sec.~\ref{app:programs}). Six cube blocks give a quartic CCZ factory with \QuarticUB\ T inputs. Only 54 of the cost-four combinations are harmful. Each consists of two faulty blocks with given errors and arises from $4\times4$ choices of input fault pairs. Hence $\epsilon=\QuarticCoef p^4+O(p^5)$, less than a third of the previous best leading error at 48 inputs. Replacing each cube block by its eight T columns and one check row turns the design into a generalized triorthogonal matrix with ten independent check rows, so the improvement lies within that family.

\paragraph*{A CCCZ state from six T states.}
Projecting $|T\rangle^{\otimes6}$ onto the $[[6,4,2]]$ code with $Z^{\otimes6}=+1$ and $X^{\otimes6}=-1$ yields $X^{\otimes4}|\CCCZ\rangle$ up to Clifford phases, with probability $1/4$ (Proposition~\ref{prop:cccz}), so with Proposition~\ref{thm:ccz}, $L(\CCCZ)=6$.

\section{Discussion}
Lower bounds on distillation have so far either held for all protocols but depended only weakly on $d$, or resolved $d$ but only within a family of codes. For small factories, the single-code reduction removes this trade-off: every protocol is accounted for by one code, so a search over codes covers all protocols. This settles quadratic distillation of T, CS and CCZ: adaptivity, postselection and multiple stages cannot make these factories smaller. The search is also constructive: it found the nine-input T factory, which lies outside every structured family we modeled. Complementary faults make the cubic searches feasible, and the analytic bounds, together with Lemma~\ref{lem:trade}, extend the lower bounds to every order. Since the proof of Theorem~\ref{thm:code} does not use the form of the inputs, the same reduction applies to factories fed by other noisy magic states.

Because cultivated T states~\cite{cultivation,googlecultivation} already have small errors, quadratic suppression is the practically relevant regime, and it is the one our results settle. They complement exact concentration without noise~\cite{rizzoleone} and asymptotic constructions with constant overhead~\cite{whyamasaki,golowich,nguyen}, which determine the scaling but not the small factories used in practice. Our bounds count noisy inputs per run under ideal Clifford operations, not the space-time volume of a factory; per accepted output, the nine-input T factory consumes more inputs than the ten-input protocol.

The main gaps are cubic and quartic distillation of one T state (10 to 15 and 11 to 48 inputs), cubic CS (9 to 35) and cubic and quartic CCZ (10 to 36 and 11 to \QuarticUB). Three directions look promising for closing them. All 137 collapsing CCZ codes of X-rank at most four at $n=8$, $d=2$ contain $Z^{\otimes n}$ in their stabilizer group; a proof that this holds in general would restrict the search to such codes. Stronger convex relaxations, of the protocols or of the polynomial collapse identities, could raise the constant two, which our relaxation does not improve for CS and CCZ. Finally, the two-level model, enlarged with further inner factories and with outputs taken up to Clifford equivalence, may yield smaller factories.

\section*{Acknowledgments}
This work was carried out with the assistance of the language model Claude Opus 5.5. The final content, analysis and conclusions remain the sole responsibility of the authors. Computations used Gurobi~12 and CaDiCaL via PySAT.

This work has been supported by the European Union's HORIZON--MSCA-2023-DN-JD programme under the Horizon Europe (HORIZON) Marie Sk\l{}odowska-Curie Actions, grant agreement 101120296 (TENORS).

\section*{Data availability}
The search code, exact verifiers, solver models, logs and certificates are available at \url{https://github.com/llorebaga/Bounds_magic}.

\bibliography{refs}

\clearpage
\setcounter{secnumdepth}{2}\setcounter{section}{0}\setcounter{equation}{0}\setcounter{table}{0}\setcounter{figure}{0}
\renewcommand{\thesection}{S\arabic{section}}
\renewcommand{\theequation}{S\arabic{equation}}
\renewcommand{\thetable}{S\Roman{table}}
\renewcommand{\thefigure}{S\arabic{figure}}
\onecolumngrid
\begin{center}\textbf{\large Supplemental Material}\end{center}

\paragraph*{Overview of methods.}
Table~\ref{tab:methods} summarizes the methods. Each entry of Table~\ref{tab:final} is either a lower bound, proved analytically, by exhaustive search or from a search by Lemma~\ref{lem:trade}, that holds for all stabilizer protocols, or the input count of an explicit factory verified in exact arithmetic. Integer and satisfiability programs are used only to find factories. When they report infeasibility, the statement concerns a restricted family of designs, and no such statement is used in Table~\ref{tab:final}.

\begin{table}[h]
\caption{Summary of the methods. LB, lower bound valid for all stabilizer protocols; UB, upper bound given by an explicit factory.}
\label{tab:methods}
\begin{ruledtabular}
\begin{tabular}{p{0.17\textwidth}p{0.30\textwidth}p{0.29\textwidth}l}
Method & Scope & Result & Reference \\
\hline
\RR Single-code reduction & \RR all stabilizer protocols, including adaptive and multistage ones & \RR a protocol exists if and only if some code satisfies the collapse conditions & Thm.~\ref{thm:code}, Sec.~\ref{app:proofs} \\
\RR Complementary faults & \RR targets with rational Pauli expectations & \RR $e$ is harmless if and only if $e+\mathbf1$ is; larger tolerated sets $H$ & Thm.~\ref{thm:pairing}, Sec.~\ref{app:proofs} \\
\RR Analytic witnesses & \RR input states that contain only harmless fault patterns & \RR LB at every order ($2d+1$, $d+\nu+1$) & Thm.~\ref{thm:analytic}, Sec.~\ref{app:analytic} \\
\RR Stabilizer-cone relaxation & \RR all protocols, through their Choi operators, twirled to T-basis diagonal form & \RR LB $n\geq2d$ for one T state, via its certificate & Thm.~\ref{thm:analytic}, Sec.~\ref{app:analytic} \\
\RR Exhaustive search & \RR all stabilizer codes on $n$ inputs, up to permutations of the inputs and the symmetry $W$ & \RR LB when no code satisfies the collapse conditions; a factory when one does (T, $n=9$) & Propositions~\ref{thm:tcs},~\ref{thm:ccz}, Sec.~\ref{app:search} \\
\RR Trading order for inputs & \RR all protocols, by feeding some inputs with maximally mixed qubits & \RR LB at order $d$ from an LB at order $s<d$ & Lemma~\ref{lem:trade}, Sec.~\ref{app:proofs} \\
\RR Clifford orbits & \RR Clifford-equivalent forms of the target & \RR equivalent targets of lower T-count ($\CS_{12}\CS_{13}\simeq\CCZ$) & Prop.~\ref{prop:orbit} \\
\RR Integer program & \RR binary matrices $G$ with parity constraints & \RR finds single-level factories (rediscovers the 36-input code); no bound depends on it & Sec.~\ref{app:programs} \\
\RR SAT & \RR T columns and CCZ blocks of two-level designs & \RR finds multistage factories (UB) & Sec.~\ref{app:programs} \\
\RR Exact certification & \RR each explicit construction & \RR collapse conditions, error polynomial and output state, computed independently of the solvers & Secs.~\ref{app:search},~\ref{app:programs} \\
\end{tabular}
\end{ruledtabular}
\end{table}

\section{A worked example}\label{app:example}
We illustrate the method on a protocol with three inputs, and then apply the same test to a seven-input code that is excluded by the exhaustive search and to the eight-input CCZ factory.

\paragraph*{Step 1: the protocol.}
Take three noisy T states. Measure $Z_1Z_2Z_3$ and keep the outcome $+1$. Then apply CNOT gates from qubits 1 and 2 onto qubit 3, which leaves qubit 3 in $|0\rangle$, and discard it. We show below that the output is $\CS|{+}{+}\rangle$ up to Clifford gates, with $\CS=\mathrm{diag}(1,1,1,i)$.

\paragraph*{Step 2: the code.}
In the notation of Theorem~\ref{thm:code}, the protocol projects onto the code $C$ fixed by $Z_1Z_2Z_3$, with projector $\Pi_C=(I+Z_1Z_2Z_3)/2$. It is spanned by the even-weight strings $|000\rangle,|011\rangle,|101\rangle,|110\rangle$ and has dimension $4=2^r$ with $r=2$. The decoder is the inverse of the Clifford encoding $U|y_1y_2\rangle=|y_1,y_2,y_1\oplus y_2\rangle$.

\paragraph*{Step 3: the fault-free input.}
Since $|T\rangle^{\otimes3}=2^{-3/2}\sum_x\omega^{|x|}|x\rangle$ and $\omega^2=i$, the projection keeps the even strings,
\begin{equation}
 \Pi_C|\phi_{\mathbf0}\rangle=2^{-3/2}\bigl(|000\rangle+i|011\rangle+i|101\rangle+i|110\rangle\bigr).
\end{equation}
Decoded, the amplitudes on $|y_1y_2\rangle$ are $(1,i,i,i)/2^{3/2}$, that is, $i^{y_1+y_2}(-i)^{y_1y_2}$ up to normalization. This is $(S\otimes S)\,\CS^{-1}|{+}{+}\rangle$, and $\CS^{-1}=\CS\cdot\mathrm{CZ}$, so the output is $\CS|{+}{+}\rangle$ up to Clifford gates. The Gram entry $G_{\mathbf0\mathbf0}=\|\Pi_C\phi_{\mathbf0}\|^2=1/2$ is the probability of accepting the ideal input.

\paragraph*{Step 4: the faulty inputs.}
Every $Z^e$ commutes with $Z_1Z_2Z_3$, so $\Pi_C|\phi_e\rangle=Z^e\,\Pi_C|\phi_{\mathbf0}\rangle$, and $G_{ee}=1/2$ for all $e$. Two cases occur.
\begin{itemize}
\item $e=111$. Here $Z^e=Z_1Z_2Z_3$ is the stabilizer itself, so $\Pi_C|\phi_{111}\rangle=\Pi_C|\phi_{\mathbf0}\rangle$: the all-faulty input collapses onto the fault-free one. In Gram form, $G_{\mathbf0\mathbf0}G_{ee}-|G_{\mathbf0e}|^2=\tfrac12\cdot\tfrac12-\tfrac14=0$. This is the cross condition of Theorem~\ref{thm:pairing} in its simplest form.
\item $e=100$, and likewise every pattern of weight one or two. Now $Z^e$ commutes with the stabilizer without being one, so it acts on $C$ as a logical Pauli operator; for $e=100$ it is $Z$ on the first logical qubit. The output becomes $Z_1\CS|{+}{+}\rangle$, which differs from the target. Indeed $G_{\mathbf0e}=\tfrac18(1+1-1-1)=0$, so the minor is $\tfrac14\neq0$. Lemma~\ref{lem:logical} predicts this without computation.
\end{itemize}

\paragraph*{Step 5: the error.}
Every pattern is accepted with probability $1/2$, and only $000$ and $111$ are harmless, so
\begin{equation}
 \epsilon(p)=1-(1-p)^3-p^3=3p(1-p).
\end{equation}
The suppression order is one. The error is invariant under $p\mapsto1-p$ because $Z^{\otimes3}$ is a stabilizer, which is stronger than what Theorem~\ref{thm:pairing} requires. Theorem~\ref{thm:analytic} shows that suppression order two needs at least five inputs, and Proposition~\ref{thm:tcs} raises this to eight.

\paragraph*{Step 6: a code rejected by the search.}
For quadratic CCZ with seven inputs, the tolerated set $H$ contains the fault-free pattern, the seven single faults, their seven complements and the all-faulty pattern. The first code of X-rank one that is not excluded by Lemma~\ref{lem:logical} has stabilizers
\begin{equation}
 X_1X_2X_3X_4X_5X_6,\quad Z_7,\quad Z_1Z_6,\quad Z_2Z_3Z_4Z_5.
\end{equation}
Lemma~\ref{lem:logical} does not exclude it, because every tolerated $Z^e$ that commutes with all stabilizers is itself a stabilizer: $Z_7$, $Z_1\cdots Z_6=(Z_1Z_6)(Z_2Z_3Z_4Z_5)$ and $Z^{\otimes7}$. Every other single fault is detected. With all signs $+1$, the exact Gram entries are $G_{\mathbf0\mathbf0}=1/16$ and, for a fault on input 1, $G_{\mathbf0e}=-i/32$ and $G_{ee}=1/16$. Hence
\begin{equation}
 G_{\mathbf0\mathbf0}G_{ee}-|G_{\mathbf0e}|^2=\tfrac1{256}-\tfrac1{1024}=\tfrac3{1024}\neq0 .
\end{equation}
The projection of the faulty input has a component orthogonal to that of the ideal input, so the output state is wrong. Each of the 16 choices of signs violates twelve of the fifteen conditions. The search applies this test to all 7{,}699{,}696 signed codes not excluded by Lemma~\ref{lem:logical}, and none satisfies it.

\paragraph*{Step 7: the eight-input factory passes.}
Label the eight inputs by the vertices $v\in\F^3$ of a cube. Writing the T$^\dagger$ inputs of the usual construction as $S^\dagger|T\rangle$, the stabilizers are $X$ on the four even-weight vertices times $Y$ on the four odd-weight vertices, together with $Z$ on the faces $v_1=0$, $v_2=0$, $v_3=0$ and $v_1=1$. These generate $Z$ on all six faces and $Z^{\otimes8}$. The collapse test gives the following.
\begin{itemize}
\item $G_{\mathbf0\mathbf0}=1/16$, and $\Pi_C\phi_{\mathbf0}$ has the Pauli spectrum of CCZ, that is, the same multiset of absolute Pauli expectations~\cite{beverland}.
\item Every single fault anticommutes with the first stabilizer, and indeed $G_{ee}=G_{\mathbf0e}=0$: single faults are always rejected.
\item Since $Z^{\otimes8}$ is a stabilizer, each pattern behaves like its complement. The weight-seven patterns are rejected, and the all-faulty input collapses onto the fault-free one, with $G_{\mathbf0\mathbf1}=G_{\mathbf1\mathbf1}=1/16$.
\item Every $Z_jZ_k$ commutes with all stabilizers but is not one, so it is a logical error. All 28 weight-two patterns are harmful, and $\epsilon=28p^2+O(p^3)$.
\end{itemize}
Eight inputs thus achieve order two, whereas, by Proposition~\ref{thm:ccz}, seven do not.

\section{Proofs of the reductions}\label{app:proofs}
\begin{proof}[Proof of Theorem~\ref{thm:code}]
Write the accepted map as $\Lambda=\sum_bc_b\Lambda_b$ over branches, where a branch fixes all random choices and measurement outcomes. Each branch damage $h_{b,e}=\tr[(I-\Pi)\Lambda_b(\rho_e)]$ is nonnegative, and these damages sum to the damage of $e$, so every branch is harmless on $H$. Measuring each discarded qubit in the computational basis refines the branches into branches given by single Kraus operators $K$. Pick one with $K|\phi_{\mathbf0}\rangle\neq0$, which exists because the fault-free input is accepted. $K$ is a product of Clifford gates, stabilizer preparations and Pauli projections~\cite{campbellbrowne}, so applied to half of $n$ Bell pairs it yields a pure stabilizer state on $n+r$ qubits. Every bipartite pure stabilizer state equals, up to local Clifford gates, a product of Bell pairs and single-qubit states~\cite{fattal}. Hence $K=c\,V\Pi_C$, where $C$ is a stabilizer code on the inputs and $V$ a Clifford isometry from $C$ onto its range. Since $K|\phi_{\mathbf0}\rangle\propto|\tau\rangle$ and $\tau$ is fixed by no Pauli operator other than the identity, the range is the whole output space, so $\dim C=2^r$ and $U=V^{-1}$ is an encoding. Harmlessness of $e$ in this branch is Eq.~\eqref{eq:collapse}. Conversely, measuring the stabilizers of $C$, postselecting the prescribed signs and applying $V$ is a protocol. For the Gram form, $G_{\mathbf0\mathbf0}>0$ is $\Pi_C|\phi_{\mathbf0}\rangle\neq0$, and given this, $\Pi_C|\phi_e\rangle\in\mathbb C\,\Pi_C|\phi_{\mathbf0}\rangle$ (including $\Pi_C|\phi_e\rangle=0$) is equality in the Cauchy--Schwarz inequality $|G_{\mathbf0e}|^2\leq G_{\mathbf0\mathbf0}G_{ee}$.
\end{proof}

\begin{proof}[Proof of Theorem~\ref{thm:pairing}]
Let $\sigma$ flip the sign of $\sqrt2$ in every Pauli coefficient of the form $x+y\sqrt2$ with $x,y$ rational. Each step of a branch maps Pauli coefficients by rational matrices (compare the closure of number fields under stabilizer circuits~\cite[Theorem~2.8]{beverland}). Cliffords permute Paulis up to sign, stabilizer preparations have rational coefficients, partial traces multiply by integers, and a measurement outcome acts as $O\mapsto(I\pm P)O(I\pm P)/4$. Hence $\sigma\circ\Lambda_b=\Lambda_b\circ\sigma$. By Eq.~\eqref{eq:input}, $\sigma(\rho_T(p))=\rho_T(1-p)$ as polynomials in $p$, so $\sigma(\rho_e)=\rho_{e+\mathbf1}$, and $\sigma$ fixes the projector onto a rational $\tau$. Thus $\sigma(h_{b,e})=h_{b,e+\mathbf1}$, and $\sigma$ maps only zero to zero. For the cross condition of the main text, extend $\sigma$ to $\mathbb Q(\omega)$ by $\sigma(i)=i$, so that $\sigma(\omega)=-\omega$ and $\sigma(|T\rangle)=-Z|T\rangle$, hence $\sigma(|\phi_e\rangle)=(-1)^n|\phi_{e+\mathbf1}\rangle$ entrywise. Since $\Pi_C$ has entries in $\mathbb Q(i)$, $\sigma(\Pi_C|\phi_{\mathbf0}\rangle)=(-1)^n\Pi_C|\phi_{\mathbf1}\rangle$, which is nonzero. As $\mathbf1$ is harmless, Eq.~\eqref{eq:collapse} makes this vector proportional to $\Pi_C|\phi_{\mathbf0}\rangle$.
\end{proof}

\begin{lemma}[Tolerated Z errors]\label{lem:logical}
In Theorem~\ref{thm:code}, if $e\in H$ is nonzero and $Z^e$ commutes with all stabilizers of $C$, then $\pm Z^e$ is a stabilizer.
\end{lemma}
\begin{proof}
Otherwise $Z^e$ is a nontrivial logical operator with $\Pi_CZ^e\phi_{\mathbf0}=Z^e\Pi_C\phi_{\mathbf0}\propto\Pi_C\phi_{\mathbf0}$, so the encoded target would be fixed by a Pauli operator.
\end{proof}

\begin{lemma}[Post-processing]\label{lem:post}
If a stabilizer branch maps $|\tau\rangle$ to $|\tau'\rangle$ with nonzero probability, then $n_{\min}(\tau',d)\leq n_{\min}(\tau,d)$. This holds in particular for Clifford equivalence and for discarding a tensor factor.
\end{lemma}
\begin{proof}
Patterns that produced exactly $\tau$ now produce exactly $\tau'$, and the fault-free input is still accepted.
\end{proof}

\begin{proof}[Proof of Proposition~\ref{prop:orbit}]
Direct computation of the eight amplitudes, checked in exact arithmetic by the supplementary code. A breadth-first search shows that the Clifford orbit of $|\CCZ\rangle$ has 69{,}120 elements up to phase. Its diagonal members of the third level of the Clifford hierarchy~\cite{gottesmanchuang} have no T terms. Up to S and CZ gates, their phase polynomials are $4x_1x_2x_3$ plus zero, one or three CS terms, or exactly two CS terms and no cubic term. These are the CNOT+S classes of CCZ and $\CS_{12}\CS_{13}$~\cite{wjs}.
\end{proof}

\begin{proposition}[{\cite[Proposition~A.4]{beverland}}]\label{prop:cs}
Projecting the third qubit of $|\CCZ\rangle$ onto $(|0\rangle\pm i|1\rangle)/\sqrt2$ leaves $\CS^{\pm1}|{+}{+}\rangle$, so $n_{\min}(\CS,d)\leq n_{\min}(\CCZ,d)$.
\end{proposition}
\begin{proof}
The remaining amplitudes are $1\mp i(-1)^{x_1x_2}$, whose ratio between $x_1x_2=1$ and $x_1x_2=0$ is $\pm i$. Since $\CS^{-1}=\CS\cdot\mathrm{CZ}$, both outcomes give $\CS|{+}{+}\rangle$ up to a Clifford gate; apply Lemma~\ref{lem:post}.
\end{proof}

\begin{proposition}\label{prop:cccz}
$L(\CCCZ)=6$.
\end{proposition}
\begin{proof}
The search excludes every code on five inputs. For the matching construction, project $|T\rangle^{\otimes6}$ onto $Z^{\otimes6}=+1$ and $X^{\otimes6}=-1$; this succeeds with probability $1/4$. The code space is spanned by $|x\rangle-|\bar x\rangle$ for even-weight $x$. On the representatives $x=(y,y_1\oplus\dots\oplus y_4,0)$, $y\in\F^4$, the amplitudes are proportional to $(-1)^{e_3(y)+e_4(y)}$, where $e_k$ are the elementary symmetric polynomials. Since $e_4(y\oplus\mathbf1)=e_4+e_3+e_2+e_1+1$ modulo two, this is $X^{\otimes4}|\CCCZ\rangle$ up to the Clifford phases $(-1)^{e_1+e_2+1}$. The exclusion of five inputs also follows analytically. The code on five inputs has a single generator $sP$, and the fault-free and all-faulty inputs must collapse together (Theorem~\ref{thm:pairing}). With $a=\langle\phi_{\mathbf0}|P|\phi_{\mathbf0}\rangle$, $b=\langle\phi_{\mathbf1}|P|\phi_{\mathbf1}\rangle$ and $c=\langle\phi_{\mathbf0}|P|\phi_{\mathbf1}\rangle$, the condition reads $(1+sa)(1+sb)=|c|^2$, which a direct case check shows to hold only for $P=Z^{\otimes5}$. The parity projection of $|T\rangle^{\otimes5}$ has absolute Pauli expectations in $\{0,\tfrac14,\tfrac12,1\}$, whereas $|\CCCZ\rangle$ has $\langle X_1\rangle=\tfrac34$, so the two are not Clifford-equivalent up to stabilizer factors.
\end{proof}

\begin{lemma}[Clean acceptance]\label{lem:clean}
If a protocol with $\Lambda$ independent of $p$ has $\epsilon=O(p^d)$, then some protocol on the same inputs accepts the fault-free pattern and makes every pattern with $|e|<d$ harmless.
\end{lemma}
\begin{proof}
Let $a_e=\tr\Lambda(\rho_e)$, $k=\min\{|e|:a_e>0\}$ and $\ell=\min\{|e|:h_e>0\}$, with $\ell=\infty$ if no pattern is harmful. The numerator and denominator of $\epsilon$ are $\Theta(p^\ell)$ and $\Theta(p^k)$, so $\ell\geq k+d$. Choose $e_*$ with $|e_*|=k$ and $a_{e_*}>0$, and apply $Z^{e_*}$ to the inputs first. Pattern $g$ of the new protocol is pattern $e_*+g$ of the old one, so the fault-free pattern is accepted, and every $g$ with $|g|<d$ has $|e_*+g|<k+d\leq\ell$ and is harmless.
\end{proof}

\begin{lemma}[Trading order for inputs]\label{lem:trade}
Let $\tau$ not be a stabilizer state. For $1\leq s\leq d$, $n_{\min}(\tau,d)\geq n_{\min}(\tau,s)+d-s$.
\end{lemma}
\begin{proof}
Take an order-$d$ protocol on $n$ inputs that accepts the fault-free pattern (Lemma~\ref{lem:clean}). Since $\tau$ is not a stabilizer state, $n\geq d$ (Theorem~\ref{thm:analytic}), so we can feed $k=d-s$ of its inputs with the stabilizer state $I/2=\tfrac12(|T\rangle\langle T|+Z|T\rangle\langle T|Z)$. A pattern $g$ on the remaining $n-k$ inputs then yields a uniform mixture of the original patterns $(g,f)$, $f\in\F^k$. If $|g|<s$, all of them have fewer than $s+k=d$ faults and are harmless, so the output is a multiple of the target, and for $g=\mathbf0$ the mixture contains the accepted fault-free pattern. This is an order-$s$ protocol on $n-k$ noisy inputs.
\end{proof}

\section{Analytic bounds}\label{app:analytic}
The Clifford $W=(X+Y)/\sqrt2$ fixes $|T\rangle$ and negates $Z|T\rangle$, so $W^a\phi_e=(-1)^{a\cdot e}\phi_e$. Applying $W$ to each qubit of $|\psi\rangle$ independently with probability one half removes all coherence between patterns:
\begin{equation}
 2^{-n}\textstyle\sum_aW^a|\psi\rangle\langle\psi|W^a=\sum_eq_e\rho_e,\qquad q_e=|\langle\phi_e|\psi\rangle|^2.
 \label{eq:twirl}
\end{equation}
We say that $\psi$ \emph{contains} the patterns with $q_e>0$.

\begin{lemma}[Witness]\label{lem:witness}
If $\psi$ contains the fault-free pattern and only harmless patterns, then a stabilizer branch maps $|\psi\rangle$ to $|\tau\rangle$ with nonzero probability.
\end{lemma}
\begin{proof}
Feed the mixture~\eqref{eq:twirl} to the protocol. The output $\sum_eq_ea_e\Pi$ is a positive multiple of the target. Splitting it by twirl choice and by branch writes a multiple of a pure state as a sum of positive operators. Each of them is then a multiple of $\Pi$, and at least one is nonzero.
\end{proof}

\begin{lemma}[Nullity]\label{lem:nullity}
If a stabilizer branch maps $|\psi\rangle$ to $|\varphi\rangle$, then $\nu(\psi)\geq\nu(\varphi)$, and at equality every Pauli expectation of $\psi$ is zero or $\pm$ one of $\varphi$.
\end{lemma}
\begin{proof}
The inequality is due to Beverland \emph{et al.}~\cite[Proposition~2.3]{beverland}; we repeat the argument because the equality case uses it.
Postpone discarding to the end. Ancillas and Cliffords preserve the nullity. Consider a projection $(I+P)/2$ on a pure state $\chi$. If $\pm P\in S_\chi$, it acts trivially or annihilates $\chi$. If $P$ anticommutes with some $g\in S_\chi$, it acts as the Clifford $(P+g)/\sqrt2$. Otherwise it lowers the nullity by at least one. A pure output means the remaining qubits are in a state $\varphi\otimes\kappa$ of nullity $\nu(\varphi)+\nu(\kappa)$. At equality, no projection lowers the nullity and $\kappa$ is a stabilizer state, so $|\varphi\rangle|\kappa\rangle=C|\psi\rangle|s\rangle$ for a Clifford $C$, which maps Pauli expectations to Pauli expectations up to sign.
\end{proof}

Since $|0\rangle=(|T\rangle+Z|T\rangle)/\sqrt2$, the state $|0\rangle^{\otimes n}$ contains every pattern. For even $N$, the stabilizer state $|G_N\rangle=(|{+_i}\rangle^{\otimes N}-|{-_i}\rangle^{\otimes N})/\sqrt2$ has $\langle\phi_e|G_N\rangle=(\alpha^{N-|e|}\beta^{|e|}-\beta^{N-|e|}\alpha^{|e|})/\sqrt2$, with $\alpha=(1+\omega)/2$ and $\beta=(1-\omega)/2$. Since $|\alpha/\beta|=1+\sqrt2$, this vanishes exactly when $|e|=N/2$.

\begin{proof}[Proof of Theorem~\ref{thm:analytic}, first two claims]
If $n<d$, every pattern is harmless and $|0\rangle^{\otimes n}$ would turn a stabilizer state into $\tau$, so $n\geq d$. The state $|T\rangle^{\otimes(n-d+1)}|0\rangle^{\otimes(d-1)}$ contains only patterns of weight below $d$, so Lemma~\ref{lem:witness} gives $n-d+1\geq L(\tau)$. Let $\tau$ be rational and not a stabilizer state. Lemma~\ref{lem:nullity}, applied to a branch mapping $|T\rangle^{\otimes L}$ with $L=L(\tau)$ to $\tau$, gives $L\geq\nu(\tau)$. At equality every Pauli expectation of $|T\rangle^{\otimes L}$ would be zero or $\pm$ one of $\tau$, but $\langle X\rangle=1/\sqrt2$ is irrational; hence $L(\tau)\geq\nu(\tau)+1$ and $n\geq d+\nu(\tau)$. (Equivalently, at equality the canonical form of~\cite[Theorem~5.3]{beverland} contains no projections, and $\tau$ would be Clifford-equivalent to $|T\rangle^{\otimes L}$ up to stabilizer factors.) If $n<2d$, Theorem~\ref{thm:pairing} makes every pattern harmless, and $|0\rangle^{\otimes n}$ would turn a stabilizer state into $\tau$, which is impossible since $\nu(\tau)\geq1$. If $n=2d$, only patterns of weight $n/2$ can be harmful, and $G_n$ contains none of them.
\end{proof}

For the third claim, split the inputs into a block $A$ of $a\geq2$ qubits and a block $B$ of $d$ qubits. Set $|\beta\rangle=(1+\bar\omega)^d|{+}\rangle^{\otimes d}-(1-\bar\omega)^d|{-}\rangle^{\otimes d}$ and $|\psi\rangle\propto|T\rangle^{\otimes a}|\beta\rangle+(Z|T\rangle)^{\otimes a}Z^{\otimes d}|\beta\rangle$.
\begin{lemma}\label{lem:magicwitness}
The state $\psi$ has $\nu(\psi)\leq a$. It contains the fault-free pattern, and only patterns with fewer than $d$ or more than $a$ faults. For $d\geq2$, some Pauli expectation of $\psi$ is not dyadic.
\end{lemma}
\begin{proof}
The operators $X_jX_{j+1}$ on neighboring qubits of $B$ fix both terms, and $Z^{\otimes(a+d)}$ swaps them; these are $d$ independent stabilizers. Since $\langle T|Z|T\rangle=0$, the $A$ part of every contained pattern is all-clear or all-faulty. With $\langle T|{\pm}\rangle=(1\pm\bar\omega)/2$ and $\langle ZT|{\pm}\rangle=(1\mp\bar\omega)/2$, the all-faulty pattern of $B$ is absent from $\beta$, and the fault-free one is absent from $Z^{\otimes d}\beta$. The fault-free amplitude $[(1+\bar\omega)^{2d}-(1-\bar\omega)^{2d}]2^{-d}$ is nonzero. For $X$ on one qubit of each block, $\langle X_AX_B\rangle=[(2+\sqrt2)^d-(2-\sqrt2)^d]/\{\sqrt2[(2+\sqrt2)^d+(2-\sqrt2)^d]\}$. Writing $(1+\sqrt2)^d=P+Q\sqrt2$ with $P,Q$ positive integers, this equals $Q/P$ for even $d$ and $P/(2Q)$ for odd $d$. From $P^2-2Q^2=(-1)^d$, $P$ is odd and coprime to $Q$. For even $d\geq2$ the reduced denominator is $P\geq3$, which is odd. For odd $d\geq3$ it is $2Q$, and $Q=\sum_k\binom{d}{2k+1}2^k\equiv d\equiv1$ modulo two, with $Q\geq5$; so $Q$ is an odd factor greater than one. (At $d=1$, $P=Q=1$ and the value $1/2$ is dyadic.)
\end{proof}
\begin{proof}[Proof of Theorem~\ref{thm:analytic}, third claim]
If $\nu(\tau)\leq d$, the claim already follows from $n\geq2d+1$, so let $\nu=\nu(\tau)\geq d+1$ and suppose $n\leq d+\nu$. Since $n\geq2d+1$, the block size $a=n-d$ satisfies $3\leq d+1\leq a\leq\nu$. All patterns in $\psi$ are harmless by Theorem~\ref{thm:pairing}, so Lemma~\ref{lem:witness} maps $\psi$ to $\tau$. Lemma~\ref{lem:nullity} then forces $\nu(\psi)=\nu(\tau)$ and dyadic expectations, which contradicts Lemma~\ref{lem:magicwitness}.
\end{proof}

\paragraph*{One T state.}
For one T state the bounds above give only $n\geq d$. The last claim of Theorem~\ref{thm:analytic} was found with a linear-programming relaxation. The accepted map of any protocol has a Choi operator in the cone generated by stabilizer states on $n+1$ qubits, since each single-Kraus branch has a pure stabilizer Choi state (proof of Theorem~\ref{thm:code}); for channels this is the characterization of completely stabilizer-preserving maps~\cite[Theorem~3.1]{seddoncampbell}. Twirling inputs and output by $W$ and averaging over input permutations, the protocol becomes
\begin{equation}
 \Lambda(\rho)=\sum_e\langle\phi_e|\rho|\phi_e\rangle\bigl(\lambda_{|e|}\,|T\rangle\langle T|+\bar\lambda_{|e|}\,Z|T\rangle\langle T|Z\bigr),
 \label{eq:tform}
\end{equation}
with $\lambda,\bar\lambda\geq0$, $\lambda_0>0$ and $\bar\lambda_v=0$ for $v<d$, and its Choi operator is diagonal in the T basis. Projecting some inputs onto stabilizer states must leave the rest in the stabilizer cone of fewer qubits, which gives linear constraints on $(\lambda,\bar\lambda)$; the cones of up to five qubits are computed exactly from all stabilizer states. This relaxation excludes $n=2d-1$, with exact certificates in $\mathbb Q(\sqrt2)$ for $d\leq7$, and is feasible at $n=2d$. For CS and CCZ the same construction, with the output dephased in the basis $Z^a|\tau\rangle$, reproduces $2d+1$ but does not improve it at the levels we tried. The certificates have a simple structure, which yields the following proof.

\begin{proof}[Proof of Theorem~\ref{thm:analytic}, last claim]
By padding it suffices to exclude $n=2d-1$. By Theorem~\ref{thm:code} we may take a single branch, whose Choi operator is a stabilizer state with rational Pauli coefficients, and twirl it to the form~\eqref{eq:tform}; the twirl keeps the coefficients rational. The automorphism $\sigma$ then fixes the Choi operator, maps $\rho_e$ to $\rho_{e+\mathbf1}$ and exchanges $|T\rangle$ with $Z|T\rangle$, so $\lambda_v=0$ for $v>n-d$, that is, for $v\geq d$.

Keep one input and the output, and project the other $2d-2$ inputs onto a stabilizer state $\chi$ with T-basis weight distribution $\mu_v=\sum_{|e|=v}|\langle\phi_e|\chi\rangle|^2$. The result is a two-qubit operator in the stabilizer cone, diagonal in the T basis, with entries $g_{00}=\sum_v\mu_v\lambda_v$, $g_{01}=\sum_v\mu_v\bar\lambda_v$, $g_{10}=\sum_v\mu_v\lambda_{v+1}$ and $g_{11}=\sum_v\mu_v\bar\lambda_{v+1}$ (first index for the kept input, second for the output; the transpose in the Choi operator replaces $|T\rangle$ by $S^\dagger|T\rangle$ on the kept input, which an $S$ gate absorbs). With $\eta=3-2\sqrt2=\tan^2(\pi/8)$, every element of the cone satisfies
\begin{equation}
 g_{01}-\eta^2g_{10}\geq\eta\,|g_{00}-g_{11}|,\qquad g_{10}-\eta^2g_{01}\geq\eta\,|g_{00}-g_{11}|.
 \label{eq:twoqubit}
\end{equation}
By the triangle inequality it suffices to check pure two-qubit stabilizer states, which are products or maximally entangled. For a maximally entangled state the entries are $|M_{ij}|^2/2$ for a unitary $M$, so $g_{00}=g_{11}$ and $g_{01}=g_{10}\geq0$, and both inequalities hold. For a product state, $g_{ij}=\tfrac14(1+(-1)^iu)(1+(-1)^jv)$, where $u,v\in\{0,\pm1/\sqrt2\}$ are the expectations of $W$ on the two factors, and a factor with expectation 0 is the average of those with $\pm1/\sqrt2$. With $c_\pm=(1\pm1/\sqrt2)/2$, so that $c_-=\eta c_+$, equal signs give $g_{01}=g_{10}=c_+c_-$ and $|g_{00}-g_{11}|=c_+^2-c_-^2$, with equality in Eq.~\eqref{eq:twoqubit} since $c_+c_-(1-\eta^2)=\eta(c_+^2-c_-^2)$, and opposite signs give $g_{00}=g_{11}$ and $\{g_{01},g_{10}\}=\{c_+^2,\eta^2c_+^2\}$, so both inequalities hold in every case. A check over all 60 two-qubit stabilizer states confirms this.

The singlet carries exactly one fault in the T basis ($\mu=\delta_1$), whereas $|{+}\rangle$ and $|{-}\rangle$ have $\mu\propto(1,\eta)$ and $\mu\propto(\eta,1)$. For $t=0,\dots,d-1$, let $\chi_t^\pm$ consist of $d-1-t$ singlets and $|{\pm}\rangle^{\otimes2t}$, so that $\mu_{d-1-t+u}\propto\binom{2t}{u}\eta^{\pm u}$. We show by induction on $t$ that $\lambda_{d-1}=\bar\lambda_d$ and $\lambda_{d-1-t}=\bar\lambda_{d+t}=0$ for $1\leq t\leq d-1$. For $t=0$, $\chi_0$ consists of singlets only, so $g_{01}=g_{10}=0$, and the first inequality of~\eqref{eq:twoqubit} gives $\lambda_{d-1}=g_{00}=g_{11}=\bar\lambda_d$. For $t\geq1$, the induction hypothesis leaves, for $\chi_t^+$, $g_{01}-\eta^2g_{10}=\binom{2t}{t+1}\eta^{t+1}(\bar\lambda_d-\lambda_{d-1})=0$ and $g_{00}-g_{11}=\lambda_{d-1-t}-\eta^{2t}\bar\lambda_{d+t}$, so the first inequality gives $\lambda_{d-1-t}=\eta^{2t}\bar\lambda_{d+t}$. For $\chi_t^-$ the second inequality gives in the same way $\bar\lambda_{d+t}=\eta^{2t}\lambda_{d-1-t}$. Since $0<\eta<1$, both vanish. At $t=d-1$ this is $\lambda_0=0$, a contradiction.
\end{proof}

For CCZ these bounds give six inputs at $d=2$, whereas $n_{\min}(\CCZ,2)=8$; closing the gap requires the full collapse condition, which the exhaustive search of Sec.~\ref{app:search} tests.

\section{Exhaustive search}\label{app:search}
\paragraph*{Parametrization.}
For a code with $m=n-r$ generators, the X parts of the stabilizers form a subspace $A$ of dimension $a$, and the Z-type stabilizers $Z^b$ form a subspace $B\subseteq A^\perp$ of dimension $m-a$. A basis $x_i$ of $A$ has lifts $X^{x_i}Z^{z_i}$, unique modulo $B$, which commute if and only if $x_i\cdot z_j=x_j\cdot z_i$. A \emph{signed code} is such a code together with a choice of signs of its $m$ generators. All $2^m$ sign choices are treated at once by a Walsh--Hadamard transform over the stabilizer group. We call the codes of a given X-rank $a$ a \emph{sector} of the search.

\paragraph*{Symmetry reduction and exact arithmetic.}
$A$ is taken up to input permutations, as one canonical representative per orbit, given by its reduced echelon basis $x_1,\dots,x_a$. For each such $A$ we enumerate every $B\subseteq A^\perp$ and every lift, with no further reduction by the permutations that fix $A$. Lifts are reduced modulo $B$, so their bits at the pivots of $B$ vanish. The Clifford $W_j$ on input $j$ maps $X_j\mapsto Y_j$ and $Z_j\mapsto-Z_j$; it leaves $A$ and $B$ unchanged, adds $x_{i,j}$ to bit $j$ of every $z_i$, and, since $W^s\phi_e=(-1)^{s\cdot e}\phi_e$, preserves every $|G_{ef}|$. So for each input $j$ in the support of $A$ that is not a pivot of $B$ we fix $z_{i,j}=0$ for the first $i$ with $x_{i,j}=1$; inputs outside the support of $A$ need no such normalization. Lemma~\ref{lem:logical} removes every pair $(A,B)$ for which some nonzero $e\in H$ orthogonal to $A$ lies outside $B$. The Gram entries follow from $\langle\phi_e|i^pX^xZ^z|\phi_f\rangle=[k\subseteq x]\,2^{-|x|/2}\,i^{p+2|e\wedge x|-|k|}$, where $k=z\oplus e\oplus f$ is a sum in $\F^n$ and the exponent is an integer. After scaling they lie in $\mathbb Z[\sqrt2,i]$, and every minor is tested exactly.

\paragraph*{Fault sets.}
For rational targets (CS, CCZ, CCZ$^{\otimes2}$, CCCZ) every run takes $H=\{e:|e|<d\text{ or }|e|>n-d\}$ by Theorem~\ref{thm:pairing}, both in Lemma~\ref{lem:logical} and in the collapse test; at $d=1$ this is $\{\mathbf0,\mathbf1\}$. The runs labelled ``3-qubit, $\nu=3$'' and ``T'' in Table~\ref{tab:runs} use $H=\{e:|e|<d\}$ only.

\begin{table}[t]
\caption{Exhaustive searches underlying Propositions~\ref{thm:tcs} and~\ref{thm:ccz}. The last column gives the number of signed codes that are not excluded by Lemma~\ref{lem:logical} and for which the collapse conditions were therefore tested in exact arithmetic. In no case does a code satisfy them with an encoded state that has the Pauli spectrum of the target. In the rows CCZ and ``3-qubit, $\nu=3$'', no code satisfies the collapse conditions at all, before any test of the encoded state, so these exclusions hold for every target covered by Proposition~\ref{thm:ccz}. Except in the rows ``3-qubit, $\nu=3$'' and T, the tolerated set $H$ contains the complementary patterns $|e|>n-d$. The quartic CCZ run is a consistency check, since Lemma~\ref{lem:trade} gives the quartic bound from the cubic one.}
\label{tab:runs}
\begin{ruledtabular}
\begin{tabular}{cccr}
target & $d$ & $n$ & signed codes tested \\
\hline
T & 2 & 8 & $1.32\times10^{11}$ \\
CS & 2 & 7 & 99{,}429{,}248 \\
CS & 3 & 8 & 9{,}240{,}117{,}248 \\
3-qubit, $\nu=3$ & 2 & 7 & 17{,}161{,}440 \\
CCZ (complements) & 2 & 7 & 7{,}699{,}696 \\
CCZ & 3 & 8 & 543{,}178{,}752 \\
CCZ & 3 & 9 & $9.13\times10^{11}$ \\
CCZ & 4 & 9 & $1.39\times10^{11}$ \\
CCZ$^{\otimes2}$ & 1 & 7 & 172 \\
CCZ$^{\otimes2}$ & 2 & 9 & 3{,}678{,}216 \\
CCCZ & 1 & 5 & 44 \\
\end{tabular}
\end{ruledtabular}
\end{table}

\paragraph*{Target tests.}
The collapse conditions do not determine which state is encoded. Since the code space is $U(\mathbb C^{2^r}\otimes|0\rangle^{\otimes m})$ for a Clifford encoding $U$, the normalized $n$-qubit state $\Pi_C|\phi_{\mathbf0}\rangle/\sqrt{G_{\mathbf0\mathbf0}}$ must be Clifford-equivalent to $|\tau\rangle|0\rangle^{\otimes m}$, with the $m=n-r$ syndrome qubits in $|0\rangle$. The Python implementation retains a code that satisfies the collapse conditions only if the multiset of absolute Pauli expectations of this normalized state equals that of $|\tau\rangle|0\rangle^{\otimes m}$, a necessary condition for Clifford equivalence. The compiled kernel tests the collapse conditions only.

For one T state at $n=8$ and $d=2$, with $H=\{e:|e|<2\}$, the search covers all X-ranks, $1.32\times10^{11}$ signed codes in total. The sectors of X-rank at most four (Python, with the target test) contain no T-state protocol, and in the sector of X-rank seven ($1.2\times10^{11}$ signed codes, compiled kernel) no code satisfies the collapse conditions. In the sectors of X-rank six ($1.1\times10^{10}$ signed codes) and five ($9.1\times10^{8}$), 890{,}940 and 346{,}136 codes satisfy them, and we classified these codes exactly. For each code we compute two logical Paulis by linear algebra over $\F$ and the three logical expectations of the normalized state $\Pi_C\phi_{\mathbf0}/\sqrt{G_{\mathbf0\mathbf0}}$. A single-qubit Clifford only permutes these up to sign, and for $|T\rangle$ their moduli are $\{1/\sqrt2,1/\sqrt2,0\}$. All of these codes of X-rank six, and all but 928 of X-rank five, have one modulus equal to one and so encode stabilizer states. The remaining 928 have moduli $\{0,1/3,2\sqrt2/3\}$: they encode a non-stabilizer state, but not one in the Clifford orbit of $|T\rangle$. This classification agrees with a computation of the full Pauli spectrum on random samples and on all stored codes that encode non-stabilizer states, and it correctly identifies a test code that encodes $|T\rangle$.

At $n=9$ we ran the compiled kernel on the sectors of X-rank at most six ($2.2\times10^{11}$ signed codes) and classified in the same way the $2.5\times10^{7}$ codes that satisfy the collapse conditions. Almost all encode stabilizer states, a few thousand encode other non-stabilizer states, and 96 encode a state whose logical Pauli expectations have the moduli of those of $|T\rangle$. These 96 codes all accept the ideal input with probability $1/128$ and have the same error polynomial, and they are equivalent to the code of Proposition~\ref{prop:t9}. For that code the logical operators $\bar X=-\mathrm{IIIIZIZXI}$, $\bar Y=-\mathrm{IIIXIXYII}$ and $\bar Z=-\mathrm{IIIXZXXXI}$ commute with all stabilizers and satisfy $\bar X\bar Y=i\bar Z$ on the code, and the projected ideal input has expectations $(1/\sqrt2,1/\sqrt2,0)$, those of $|T\rangle$. The acceptance probability and the error are $A(p)=D(p)/256$ and $\epsilon(p)=N(p)/D(p)$ with $D(p)=2-15p+87p^2-256p^3+408p^4-336p^5+112p^6$ and $N(p)=27p^2-141p^3+324p^4-384p^5+224p^6-48p^7$. An independent certificate recomputes these quantities and all single-fault minors in exact arithmetic over $\mathbb Q(\sqrt2,i)$; the harmful patterns number 36, 45, 90, 75, 76, 15, 6 and 1 at weights two to nine, and $\epsilon=\tfrac{27}{2}p^2+O(p^3)$. A computation with dense state vectors confirms the collapse conditions and the Pauli spectrum of $|T\rangle$ for all 96 codes. The sectors of X-rank seven and eight were not run, since the factory already shows that nine inputs suffice.

\paragraph*{Monotonicity in $n$ and $d$.}
A protocol on $n-1$ inputs, extended by one discarded input, keeps its suppression order, so each search at fixed $n$ also excludes all smaller $n$. By Lemma~\ref{lem:trade}, an exclusion at order $s$ with $n$ inputs also excludes $n+d-s$ inputs at order $d>s$. In particular, the cubic CS exclusion at $n=8$ follows from the quadratic one at $n=7$, the quartic CCZ bound $n\geq11$ follows from the cubic one, and the quartic run at $n=9$ is a consistency check.

\paragraph*{Validation.}
For $n\leq5$ the search agrees with brute force over all signed codes. Six hundred random codes at $n=6,7$ are mapped by the symmetries onto codes enumerated by the search, with identical Gram matrices, and the exact and floating-point Gram entries agree. The search recovers Jones' four-T CCZ preparation, the three-T CS preparation and the eight-T cube factory. A separately written implementation, without the $W$ symmetry and with its own arithmetic, confirms the CCZ runs at $n=7$. A compiled kernel treats the largest sectors and reproduces the results of the Python implementation in every smaller case. The two-CCZ and CCCZ runs use the Python implementation only. Table~\ref{tab:runs} lists the searches on which Propositions~\ref{thm:tcs} and~\ref{thm:ccz} rest.

\paragraph*{Verification levels.}
The computational results are verified at three levels. Reproduced by two independent implementations: quadratic CCZ ($n\leq7$) and cubic CCZ at $n=7$. One implementation, with the compiled kernel cross-checked against the Python search on all smaller cases: one T state at $n=8$, CS, cubic CCZ at $n=8,9$, quartic CCZ at $n=9$, two CCZ states and CCCZ. Exact certificates, recomputed by separate scripts in exact arithmetic: every construction of the main text, including the nine-input T factory, and the worked examples of Sec.~\ref{app:example}.

\section{Integer and SAT programs}\label{app:programs}
The programs of this section only search for constructions. No lower bound depends on them, and every design they return is certified independently in exact arithmetic.
\paragraph*{Phase-polynomial factories.}
We follow the construction of triorthogonal and generalized triorthogonal factories~\cite{bravyihaah,haah}. Registers $v=(y,c)$, with $k$ logical and $s$ check bits, start in $|{+}\rangle$. Each column $G_j$ applies T to the parity $G_j\cdot v$, consuming one input, so the phase is $\omega^{g(v)}$ with $g(v)=\sum_j[G_j\cdot v]$, where $[\cdot]\in\{0,1\}$ is a parity read as an integer. A faulty input $j$ applies $ZT$ instead of T to that parity, an extra sign $(-1)^{G_j\cdot v}$. These signs multiply, with no cross terms, so a fault pattern $e$ of any weight contributes $\prod_{j:e_j=1}(-1)^{G_j\cdot v}=(-1)^{(Ge)\cdot v}$. Suppose every overlap $\sum_jG_{aj}G_{bj}G_{cj}$ (rows not necessarily distinct) that involves a check row is even. Then the check-involving part $P(y,c)=g(y,c)-g(y,0)$ has even linear, doubly even quadratic and no cubic terms modulo eight, so $\omega^{-P}$ is a diagonal Clifford. Undoing it and projecting the checks onto $|{+}\rangle^{\otimes s}$ gives, on an input fault $e$,
\begin{equation}
 \mathrm{out}_e(y)=\textstyle\sum_c\omega^{g(y,c)-P(y,c)}(-1)^{(Ge)\cdot(y,c)},
 \label{eq:out}
\end{equation}
which is one branch of a stabilizer protocol.

\paragraph*{Integer program.}
The variables are the entries $G_{aj}\in\{0,1\}$, binary variables $y_{R,j}$ that linearize the products $\prod_{a\in R}G_{aj}$ over sets $R$ of at most three rows, and integer slacks that fix the parity of every overlap $\sum_jy_{R,j}$: even if $R$ contains a check row, and otherwise equal to the target's T, CS or CCZ coefficient. At distance two every column has a nonzero check part, and at distance three we also require the check parts to be pairwise distinct, which loses nothing for the least number of inputs, since two equal columns combine to a Clifford phase and can be deleted together. Sorting the columns breaks the permutation symmetry, and we solve one program for every CNOT+S class in the Clifford orbit of the target. Gurobi~12 finds the cube factory and the 15-to-1 protocol within seconds and the 36-input matrix below in 15 minutes. Within this family it also shows that no matrix with nine columns and at most six check rows gives a quadratic T factory, so the nine-input factory, which in this form would need six check rows (nine columns and two Z-type stabilizers), is not of generalized triorthogonal form. Such results concern this family only and are not used in Table~\ref{tab:final}.

\paragraph*{Certification.}
For every design we check exactly that $\omega^{-P}$ is Clifford, and evaluate Eq.~\eqref{eq:out} in $\mathbb Z[\omega]$ for every fault pattern below the target order. We test collapse through all $2\times2$ minors, count the harmful patterns at the target order, and compare the Pauli spectrum of the output with that of the target. The cube factory and 15-to-1 give $28p^2$ and $35p^3$.

\paragraph*{The 36-input cubic factory.}
For $\tau=\CS_{12}\CS_{13}$, $n=36$, $s=6$ and $d=3$, the integer program returns the matrix below (logical rows first). The exact check shows that the correction is Clifford, that all patterns of weight at most two are harmless, and that 120 of the 7140 weight-three patterns are harmful. After diagonal S and CZ corrections, the inverse of the Clifford of Proposition~\ref{prop:orbit} maps the output to $|\CCZ\rangle$. Reference~\cite{wjs} lists the same parameters and coefficient for this class.
{\footnotesize
\begin{verbatim}
111010110100001011110010000110101010
000100100111111001100001100001010000
010100001010111011010011010111100100
100101100111010100010110010111011101
101100100010110010110010001010001011
011100010001110001110001000101000111
000011110000001111110000111100111111
000000001111111111110000000011111111
000000000000000000001111111111111111
\end{verbatim}}

\paragraph*{Two-level model.}
Besides T columns, the outer code may contain CCZ blocks $(a,b,c)$ of three parity vectors. A block is a CCZ state from an inner factory, consumed by gate teleportation, and it multiplies the outer state by $(-1)^{[a\cdot v][b\cdot v][c\cdot v]}$. Modulo eight this phase is $4\,(a\cdot v)(b\cdot v)(c\cdot v)$. Its quadratic and linear parts are CZ and Z gates. Its cubic part adds to every distinct triple of rows the symmetrized tensor $\sum_\pi a_{\pi_1}b_{\pi_2}c_{\pi_3}$ modulo two. The twirl by the CCZ-preserving Cliffords $X_i\mathrm{CZ}_{jk}$ turns any error of a block into a mixture of $Z^l$ with $l\in\F^3\setminus\{0\}$, which acts on the outer state as $Z^{l_1a+l_2b+l_3c}$. A Jones block (four T, inner order one) has fault cost one; a cube block (eight T, inner order two) has fault cost two. Suppose every combination of faulty units (T columns or blocks, with any patterns) of total cost below $D$ has a nonzero check part or a zero logical part. Then fewer than $D$ input faults are always harmless, and by Theorem~\ref{thm:code} and Lemma~\ref{lem:post}, applied stage by stage, the composite has order $D$.

The conditions are the parity constraints of every overlap, now including the cubic contributions of the blocks, and one constraint for each \emph{fault configuration}: a set of T columns and blocks of total cost below $D$, with a nonzero error label $l\in\F^3$ for each block, whose Z-vector must have a nonzero check part or a zero logical part. The total number of T inputs is fixed by the instance, so each instance is a feasibility problem. An integer-program version is correct but slow.

The SAT instances impose these constraints on Boolean variables for the T columns and the block parity vectors. Products are Tseitin variables, every parity is an XOR chain, and each overlap and fault constraint becomes a set of clauses. Degenerate T columns are handled as follows: every T column is required to be nonzero, and the columns are in strictly increasing lexicographic order. This excludes repeated columns without loss, because removing two equal T columns from a design leaves a valid design with two fewer inputs. CaDiCaL~1.9.5 solves the instances, and each satisfying design is certified exactly as above. The six-cube-block quartic factory has outer registers $3+5$; each row below lists a block's parity vectors $(a,b,c)$, with bits in register order:
{\footnotesize
\begin{verbatim}
01111110 00001111 10100111
11101000 10110111 11111001
10101111 11011110 10110110
11000001 11100111 10010111
00010000 10110001 11110110
11010111 00010001 01011001
\end{verbatim}}
\noindent In the cube factory, every nonzero logical error pattern arises from four of its 28 harmful fault pairs, so each of the 54 harmful two-block error configurations contributes $16p^4$. Since the cube factory applies T to the eight parities $(u_1a\oplus u_2b\oplus u_3c)\cdot v\oplus c'$, $u\in\F^3$, with one check bit $c'$, each block can be replaced by these eight columns. The resulting $14\times48$ matrix has rank 13, with three logical and ten independent check rows, satisfies all overlap conditions, and has 0, 0, 0 and 864 undetected logical errors at weights one to four. As tests, the solver finds the eight-T cube factory, proves that seven T states are insufficient, and shows that a single Jones block gives CCZ at order one. For quartic CCZ, two cube blocks (three check rows) and three cube blocks (three and four check rows) are unsatisfiable. Six cube blocks are satisfiable with four to seven check rows; each design is certified, with 54 to 98 harmful two-block configurations. Five cube blocks, and four cube blocks with two Jones blocks, are not decided within 15 minutes per instance. These negative results concern the two-level model only and are not used in Table~\ref{tab:final}.

\section{Previous lower bounds}\label{app:prior}
\paragraph*{Valid under arbitrary postselection.}
Fang and Liu~\cite{fangliu,fangliu2} give $n\geq\log[(1-f_\tau)/\epsilon]/\log[(1+R)/\lambda_{\min}]$, with stabilizer fidelity $f_\tau$ of the target, generalized robustness $R$ of the input (the least $s$ such that $(\rho+s\sigma)/(1+s)$ is a stabilizer mixture for some state $\sigma$) and smallest input eigenvalue $\lambda_{\min}$. Their statement also contains a success-probability factor, which can be removed~\cite{regulaprl}. For $\rho_T(p)$, $\lambda_{\min}=p$, and by convexity $R(\rho_T(p))\leq R(|T\rangle)=3-2\sqrt2$: the Bloch vector $(1,1,0)/\sqrt2$ must be mixed with its antipode to reach the stabilizer octahedron $|x|+|y|+|z|\leq1$. Thus $1+R\leq4-2\sqrt2$, which coincides with the inverse stabilizer fidelity of $|T\rangle$, and the bound tends to $d$ as $p\to0$. (With the standard robustness, $1+R(|T\rangle)=(1+\sqrt2)/2$, the limit and the integer values below are unchanged.) At $p=10^{-3}$, with the constants of known protocols, it gives $n\geq3$ for T at $d=3$ and $n\geq2$ for CCZ at $d=2$. Regula's projective robustness~\cite{regulaprl} grows like $1/p$ and has the same limit. The nullity gives $n\geq\nu$ at every $d\geq1$, since $\epsilon(0)=0$ forces an exact postselected conversion~\cite{beverland}. Hence $n_{\min}\geq\max(d,\nu)$. For rational targets, the equality case of the canonical form of~\cite[Theorem~5.3]{beverland} raises the nullity bound to $\nu+1$ (Theorem~\ref{thm:analytic}). We have not found this bound stated in the literature, and Beverland \emph{et al.} use the automorphism $\sigma$ only to bound postselection probabilities~\cite[Proposition~5.8]{beverland}. It follows, however, immediately from their results, so Table~\ref{tab:final} marks the $d=1$ entries $\nu+1$ with a dagger rather than as new. Their stabilizer-extent bound (e.g.\ $3.63$ for CCZ) is not established under arbitrary postselection, so we do not use it; compare the probability-weighted bounds of~\cite{seddon} below.

\paragraph*{Deterministic or high-probability protocols.}
The robustness bound of Regula and Takagi~\cite{regulatakagi}, $n\geq\log[(1-\epsilon)/F]/\log(4-2\sqrt2)$, is independent of $d$. It gives $n\geq4$ for CCZ ($F=9/16$), $n\geq3$ for CS ($F=5/8$) and $n\geq8$ for two CCZ states. Seddon \emph{et al.}~\cite{seddon} weight such bounds by the success probability $P$; for CCZ they give $n\geq4$ only when $P>0.83$.

\paragraph*{Class-restricted bounds.}
Codes with a transversal T gate need at least fifteen qubits for distance three~\cite{kbk}. Generalized triorthogonal codes are classified through length 54 at distance at least three~\cite{nezamihaah,wjs}, and footprint bounds limit particular circuit families~\cite{jacinto}. None of these covers adaptive, non-transversal or multistage protocols. The nine-input T factory lies outside all of these classes. The quartic two-level design does not: flattened as in Sec.~\ref{app:programs}, it is a generalized triorthogonal code that improves the leading coefficient within that family.

\end{document}